\documentclass[symmetry,review,accept,pdftex,moreauthors]{Definitions/mdpi} 

\firstpage{1} 
\pubvolume{1}
\issuenum{1}
\articlenumber{0}
\pubyear{2024}
\copyrightyear{2024}
\externaleditor{Academic Editor: Konstantinos Gourgouliatos}
\datereceived{15 November 2024} 
\daterevised{6 December 2024} % Comment out if no revised date
\dateaccepted{9 December 2024} 
\datepublished{} 
\hreflink{\textls[-25]{https://doi.org/}} % If needed use \linebreak
\Title{Pulsar Kick: Status and Perspective}

\TitleCitation{Pulsar Kick: Status and Perspective}

\Author{Gaetano Lambiase %MDPI: Please carefully check the accuracy of names and affiliations.
$^{1,2,*}$\orcidA{} and Tanmay Kumar Poddar $^{2}$\orcidB{}} %MDPI: Please confirm whether correspondence autor is correct.

\AuthorNames{Gaetano Lambiase and Tanmay Kumar Poddar}

\AuthorCitation{Lambiase, G.; Poddar, T.K.}
\address{%
$^{1}$ \quad Dipartimento di Fisica E.R. Caianiello, Università di Salerno,
Via Giovanni Paolo II 132,\linebreak I-84084 Fisciano, (SA), Italy; lambiase@sa.infn.it\\
$^{2}$ \quad INFN, Gruppo Collegato di Salerno, Via Giovanni Paolo II 132,\linebreak I-84084 Fisciano, (SA), Italy; poddar@sa.infn.it %MDPI: Please add city, postcode and province to match the format of all affliliations
}

\corres{Correspondence: lambiase@sa.infn.it}

\abstract{The high speeds seen in rapidly rotating pulsars after supernova explosions present a longstanding puzzle in astrophysics. Numerous theories have been suggested over the years to explain this sudden "kick" imparted to the neutron star, yet each comes with its own set of challenges and limitations. Key explanations for pulsar kicks include hydrodynamic instabilities in supernovae, anisotropic neutrino emission, asymmetries in the magnetic field, binary system disruption, and physics beyond the Standard Model. Unraveling the origins of pulsar kicks not only enhances our understanding of supernova mechanisms but also opens up possibilities for exploring new physics. In this brief review, we will introduce pulsar kicks, examine the leading hypotheses, and explore future directions for this intriguing phenomenon.}

\begin{document}
%\begin{adjustwidth}
%%%%%%%%%%%%%%%%%%%%%%%%%%%%%%%%%%%%%%%%%%
%\setcounter{section}{-1} %% Remove this when starting to work on the template.
%\section{How to Use this Template}

%The template details the sections that can be used in a manuscript. Note that the order and names of article sections may differ from the requirements of the journal (e.g., the positioning of the Materials and Methods section). Please check the instructions on the authors' page of the journal to verify the correct order and names. For any questions, please contact the editorial office of the journal or support@mdpi.com. For LaTeX-related questions please contact latex@mdpi.com.%\endnote{This is an endnote.} % To use endnotes, please un-comment \printendnotes below (before References). Only journal Laws uses \footnote.

% The order of the section titles is different for some journals. Please refer to the "Instructions for Authors” on the journal homepage.
\section{Introduction}
Astrophysical compact objects like pulsars serve as remarkable cosmic markers, deepening our understanding of cosmic evolution and enabling tests of physics beyond the Standard Model (SM). Pulsars are rapidly rotating, magnetized neutron stars (NSs), first discovered by J. Bell-Burnell through observations of a radio source~\cite{Hewish:1968bj,Pilkington:1968bk}. Pulsars provide unique advantages over laboratory-based experiments, functioning as natural, highly sensitive detectors in space. Their extreme environments, expansive space as a detection medium, prolonged exposure times, and complementary role to terrestrial experiments make pulsars particularly powerful tools for studying fundamental physics at large scales. These characteristics collectively enhance interaction probabilities and sensitivity by several orders of magnitude compared to traditional laboratory setups.

Binary pulsars, in particular, represent exceptional laboratories for testing Einstein’s general relativity (GR) under intense gravitational fields. Observations of orbital period decay in binary compact star systems provided the first indirect evidence of gravitational waves (GWs), and their precise timing measurements allow for predictions with high accuracy, consistently aligned with GR~\cite{Hulse:1974eb,Kramer:2006nb}.

\textls[-15]{NSs exhibit extreme densities and pressures at their cores, reaching around $10^{18}~\mathrm{kg/m^3}$~\cite{Lattimer:2015eaa}} and approximately $10^{34}~\mathrm{Pa}$~\cite{Ozel:2016oaf}, respectively. Under such conditions, quarks and gluons may exist in a deconfined state called quark-gluon plasma, a phase thought to have existed in the early universe just microseconds after the Big Bang~\cite{Weih:2019xvw}. This state, likely unique to NSs, enables pulsars to probe conditions otherwise unattainable in the observable universe. 

A typical pulsar has a mass of approximately $1.4~M_\odot$, a radius near $10~\mathrm{km}$, and a magnetic field strength between $10^{14}-10^{16}~\mathrm{G}$~\cite{Woltjer1964}. Newly formed pulsars begin with a temperature around $\sim$10\textsuperscript{12}~{K}, which rapidly decreases to about $10^{6}~\mathrm{K}$ over a few years due to substantial energy loss through neutrino emission~\cite{Lattimer:2015eaa}. Pulsars emit electromagnetic radiation at highly regular intervals, functioning as precise cosmic clocks.

Pulsars also play a crucial role in constraining the equation of state (EoS) of nuclear matter. Measurements of pulsar mass, radius, and tidal deformability provide essential insights into the properties of ultra-dense matter. Pulsar timing arrays (PTAs), utilizing the extreme periodicity of pulsars, are used to detect the stochastic gravitational wave background (SGWB), potentially arising from supermassive black holes (SMBHs)~\cite{NANOGrav:2023gor} or exotic physics from the early universe~\cite{NANOGrav:2023hvm}.

Pulsars' strong magnetic fields create a unique platform for studying magnetohydrodynamics and plasma physics under extreme conditions. The gravitational fields of pulsars may capture dark matter (DM), altering their properties, which could affect spin-down rates and spectral lines through DM--nucleon interactions~\cite{Kouvaris:2014rja,Hook:2018iia,Fujiwara:2023hlj,Day:2019bbh,Leung:2011zz,Kain:2021hpk,Grippa:2024sfu}. Precise measurements thus serve as valuable tests for new physics scenarios, offering unique insights into the fundamental structure of the universe.

When a supergiant star reaches the end of its life, it undergoes a supernova explosion, collapsing inward and expelling its outer layers to form a gas and dust cloud known as a supernova remnant (SNR). The core collapses into an NS or pulsar, which is why most NSs and pulsars are observed near the centers of SNRs. In cases where the supernova collapse is asymmetric, the resulting pulsar can receive a "kick" in a specific direction. In~\cite{Hobbs:2005yx}, a statistical analysis of the proper motion of 233 pulsars reveals that the highest $15\%$ of the velocity distribution reaches speeds exceeding 1000 km/s—an order of magnitude greater than that of their progenitors~\cite{Long:2022qwi,Igoshev:2021bxr}. This phenomenon, known as the pulsar kick, provides a unique window into the underlying physics of supernova explosions~\cite{Burrows:1995vv}, neutrino interactions~\cite{Ayala:2019sbt}, dense matter properties~\cite{Schmitt:2005ee,Jiang:2019hal}, and even the dynamics of galaxy evolution~\cite{Chu:2021evh,Kusenko:2008gh}. Pulsar velocities can be measured using timing residuals, Doppler shifts, and parallax methods. Additionally, the expansion of associated SNRs can provide valuable insights into pulsar velocity. Observations from the Chandra X-ray Observatory and ROSAT have examined several SNRs and suggest that the asymmetry in ejecta within these remnants may provide an explanation for pulsar kicks. Additional discussions on this topic can be found at the end of Section \ref{sec2.3}.

Monte Carlo simulations estimate the average birth velocity of pulsars to be around $\sim$(250--300)~{km/s}, following a Maxwellian distribution with a dispersion of approximately $\sigma\sim$ 190~{km/s}~\cite{Hansen:1997zw}. Pulsars with longer characteristic ages exhibit an asymmetric drift, which suggests that they are dynamically older. This population may emerge from the lower-velocity segment of younger pulsars, though it is influenced by their binary origins and the effects of evolving within the Galactic potential. The findings align well with the characteristics of binaries containing NSs and closely match the velocities predicted by numerical supernova simulations. The observed velocity distribution of these pulsars is shaped by both their origins in binary systems and the influence of their later movement through the~Galaxy.

In the study by~\cite{Toscano:1998iz}, the average transverse velocity of thirteen MilliSecond Pulsars (MSPs) was determined to be around $85\pm 13~\mathrm{km/s}$, which is about four times slower than that of typical pulsars. Unlike younger ordinary pulsars, which tend to move away from the Galactic plane, half of these MSPs are moving toward it. Notably, half of them have ages comparable to or exceeding the age of the Galactic disk, indicating that they are much~older.

Johnston et al.~\cite{Johnston:2005ka} investigated the alignment between pulsar spin and velocity vectors by studying 25 pulsars at 1369 MHz, with 21 also observed at 3100 MHz. The study used precise polarization measurements, accurate rotation measures (RMs), and refined velocity data to assess the relationship between the pulsar’s rotation and velocity directions. Although this alignment is challenging to confirm due to the presence of orthogonal polarization modes, additional insights from optical and X-ray observations for specific pulsars—such as PSR B0656+14, the Crab, and Vela—help indicate that the velocity vector aligns with the rotation axis. Furthermore, many pulsars show linear polarization with the position angle predominantly perpendicular to the magnetic field lines. However, if the supernova kicks impart both linear and angular momentum to the newborn NS, it may obscure any pre-existing stellar alignment before the explosion. Thus, observations of double NS systems cannot conclusively rule out aligned kicks.

\begin{figure}[H]
%\centering
    \includegraphics[width=0.8\linewidth]{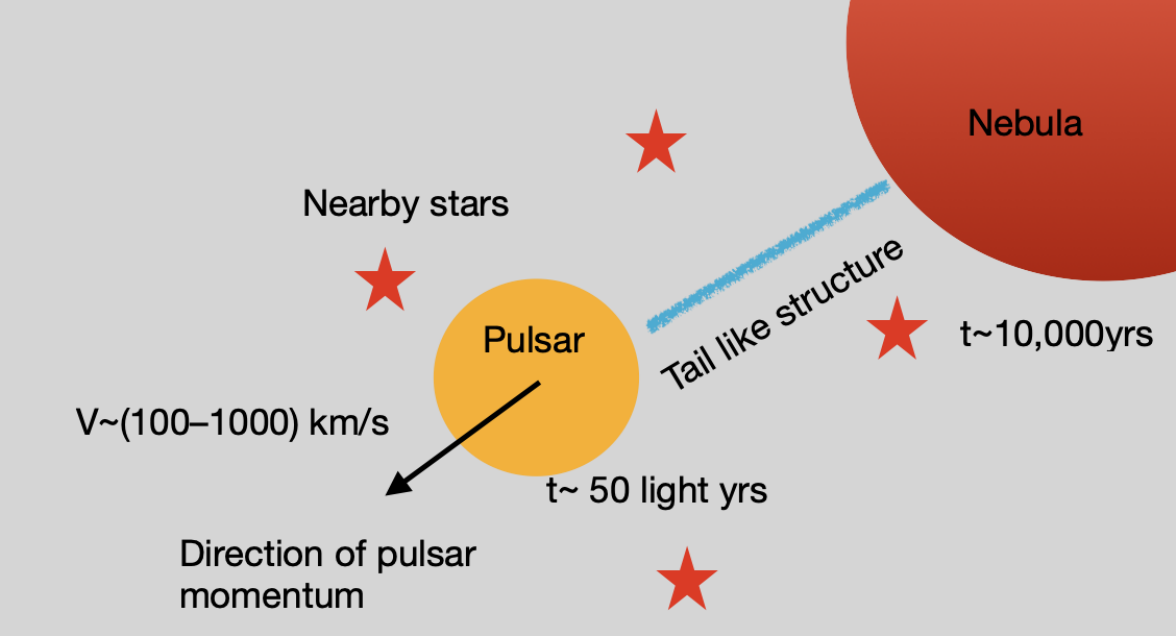}
    \caption{Schematic %MDPI: Please change the hyphen (-) into a en-dash sign (--, "U+2013"). e.g., "1-10" should be "1--10".
    %MDPI: Please check if explanation for red star should be added.
diagram of a pulsar kick. The asymmetric supernova collapse results in the pulsar a kick.}
    \label{one}
\end{figure}

In Figure \ref{one}, we present a schematic diagram illustrating a pulsar kick, where the pulsar attains a high velocity of (100--1000) {km/s} and can be observed outside the nebula. The pulsar exhibits a tail-like structure, which can be tracked using powerful radio telescopes. While, for example, the nebula is approximately 10,000 years old, the pulsar is located about $50$ light-years away.

A nice depiction of the kick of a radio pulsar, PSR J0002+6216, is shown in Figure 4 of~\cite{Schinzel:2019wlx} with false colors representing brightness temperatures ranging from $5.5~\mathrm{K}$ to a maximum of $8.9~\mathrm{K}$. The tail-like structure relative to the surrounding nebula denotes the natal kick of the pulsar, where the structure is captured in images from the very large array (VLA), with a timing analysis conducted by Fermi. The image has an angular resolution of approximately $1'$, with a field of view of about $1^\circ.9\times1^\circ.1$. A green cross marks the geometric center of the SNR~\cite{Landecker_1999}, while circles indicate the location of PSR J0002+6216~\cite{Clark:2016unw}. A faint emission tail extends from the pulsar back toward the SNR’s geometric center.

Several hypotheses have been proposed to explain these high birth velocities of pulsars. The most widely accepted explanation for pulsar kicks is anisotropy in the supernova explosion at the time of their birth~\cite{Muller:2018utr,Lai:2003hm,Nakamura:2019snn,Powell:2020cpg,Burrows:1995bb,Lai:1999hi,Nordhaus:2010ub}. However, 3D hydrodynamic simulations indicate that this mechanism struggles to produce kicks exceeding 1000 km/s~\cite{Schmitt:2005ee,Metlitski:2005pr,Muller:2018utr,Nakamura:2019snn,Powell:2020cpg}. Additionally, these models tend to predict a strong correlation between mass and velocity that is not supported by observations~\cite{Muller:2018utr,Nakamura:2019snn}. This also makes it challenging to explain low-mass NSs with nearly symmetric ejecta experiencing significant kicks~\cite{Gessner:2018ekd}. Furthermore, the models have difficulty reproducing the observed bimodal velocity distribution~\cite{Igoshev:2021bxr,Schmitt:2005ee}. While the anisotropy in supernova explosions can effectively account for low-velocity pulsars, explaining higher pulsar velocities ($\gtrsim$1000~{km/s}) presents greater challenges from a model-building perspective and necessitates additional assumptions. An additional discussion of supernova anisotropy can be found in Section \ref{sec2.3}.

Asymmetrical ejection of matter or, potentially, neutrinos during Core-Collapse SuperNovae (CCSNe) not only imparts a natal kick to newborn pulsars but also contributes to their initial spin at birth, as explored in~\cite{Fragione:2023mpe}. This study examines how these asymmetric forces create both linear and rotational motion in pulsars, shaping their velocity and spin characteristics from the moment they form.

Anisotropic neutrino emission or asymmetries in the collapsing star's magnetic field could produce such high speeds. Hydrodynamic instabilities within the supernova may impart net momentum to the NS, resulting in a kick. For progenitor stars in binary systems, the supernova explosion can disrupt the system, transferring momentum to the pulsar. Additionally, certain beyond-SM physics scenarios may contribute to the natal kick of pulsars. In the following, we provide an overview of some leading hypotheses to explain the phenomenon of pulsar kicks.

The review article is organized as follows: Section \ref{sec1} explores various mechanisms proposed to explain the phenomenon of pulsar kicks. In Section \ref{outlook}, we discuss the future perspective in the framework of pulsar kick. Finally, Section \ref{sec2} provides concluding remarks and discusses future directions for utilizing pulsar kicks as a tool to investigate fundamental physics.

We use natural units $c=1$ (the speed of light in vacuum) and $\hbar=1$ (the reduced Planck constant) in the paper unless stated otherwise.
\section{Mechanisms of Explaining Pulsar Kick}\label{sec1}
Below, we outline the leading theories proposed to explain pulsar kicks, along with their challenges and limitations. It is also possible that pulsar kicks result from a combination of all these contributing factors.
\subsection{Asymmetric Neutrino Emission}
During core collapse, a vast number of neutrinos are emitted. If this emission is slightly anisotropic, it creates a momentum imbalance that can impart a significant kick to the NS. This mechanism is particularly intriguing since even a minor asymmetry in neutrino emission can produce high velocities due to the enormous energy neutrinos carry. As a proto-NS cools, neutrinos carry away about $99\%$ of its gravitational binding energy, approximately $10^{53}\mathrm{erg}$, with a momentum of around $10^{43}\mathrm{g\cdot cm/s}$. Therefore, only a $1\%$ anisotropy in the neutrino emission distribution is sufficient to explain the pulsar kick~\cite{Lyne:1994az,Kusenko:1996sr}.

In the dense environment of an NS, electron neutrinos ($\nu_e$) have a shorter mean free path than other neutrino flavors. If resonant oscillations occur, causing muon and tau neutrinos ($\nu_{\mu,\tau}$) to oscillate into $\nu_e$ above the $\nu_{\mu,\tau}$ neutrinosphere but below the $\nu_e$ neutrinosphere, then $\nu_e$ will be absorbed by the medium due to charge and neutral current interactions. The resonance point effectively determines the $\nu_{\mu,\tau}$ neutrinosphere position. When neutrinos propagate in the presence of a magnetic field, their self-energy acquires a term proportional to $\mathbf{B}\cdot \mathbf{k}$, where $\mathbf{B}$ is the magnetic field and $\mathbf{k}$ the neutrino momentum vector. This interaction distorts the effective $\nu_{\mu,\tau}$ neutrinosphere, depending on the orientation of $\mathbf{B}$ and $\mathbf{k}$. Consequently, the neutrinosphere deviates from a spherical shape, and neutrinos escaping along the magnetic field direction have different temperatures than those emitted in the opposite direction. This directional asymmetry in neutrino momentum can create enough imbalance to account for the observed pulsar kick~\cite{Kusenko:1996sr}.

The condition for neutrino oscillation resonance in the presence of a magnetic field is given as~\cite{Kusenko:1996sr}
\begin{equation}
\frac{\Delta m^2}{2k}\cos2\theta=\sqrt{2}G_FN_e+\frac{eG_F}{\sqrt{2}}\Big(\frac{3N_e}{\pi^4}\Big)^{\frac{1}{3}}\hat{\mathbf{k}}\cdot \mathbf{B},
\label{eq:1}
\end{equation}
where $\Delta m^2$ represents the squared mass difference between neutrinos, $\theta$ is the neutrino mixing angle in a vacuum, $N_e$ is the electron number density, $G_F$ denotes the Fermi constant, $e$ is the electron charge, and $\hat{\mathbf{k}}$ is the unit vector in the direction of the neutrino momentum. In this context, the neutrino momentum asymmetry is obtained as~\cite{Kusenko:1996sr}
\begin{equation}
\frac{\delta k}{k}=1\%\Big(\frac{3~\mathrm{MeV}}{T}\Big)^2\Big(\frac{B}{3\times 10^{14}~\mathrm{G}}\Big), 
\label{eq:2}
\end{equation}
where $T$ represents the background temperature. To achieve a $1\%$ anisotropy in the neutrino momentum, the neutrino mass must be approximately $100~\mathrm{eV}$ with a small mixing angle, which should decay in the cosmological time scale not to overclose the universe. This analysis assumes that the resonance point is unaffected by the temperature distribution. 

The main limitation of this mechanism is that it requires a heavy neutrino mass, which conflicts with results from neutrino oscillation experiments and cosmological constraints~\cite{Planck:2018vyg,ParticleDataGroup:2020ssz}. Corrections to the neutrino transport and hydrostatic equations in a proto-NS indicate that a magnetic field of at least $(10^{15}-10^{16})~\mathrm{G}$~\cite{Qian:1997sw} is needed to produce a $1\%$ asymmetry in the neutrino momentum. Further refinement of the neutrino oscillation mechanism for generating pulsar kicks has considered variations in neutrino opacities and the neutrino absorption process, $\nu_e + n \rightarrow e^- + p^+$~\cite{Kusenko:1998bk}. An asymmetry in this neutrino absorption on opposite sides of the effective neutrinosphere leads to a pulsar kick, differing from the process described above, where the kick results from the distortion of the effective neutrinosphere by the magnetic field where neutrinos traveling along the magnetic field have different momenta compared to those moving in the opposite direction relative to the field. A self-consistent model of neutrino transport in the stellar atmosphere suggests that an even stronger magnetic field, approximately $10^{17}~\mathrm{G}$ near the star’s surface, is necessary to account for pulsar kicks through the oscillation mechanism~\cite{Janka:1998kb}. Later studies suggest that resonant neutrino conversion may account for pulsar kicks, with momentum asymmetry tied to the logarithmic derivative of the energy flux. If the outgoing neutrino energy flux is denoted as $F$, then the neutrino momentum asymmetry is $\delta k/k\propto h^{-1}_F$, where $h^{-1}_F=d\ln F/dr|_{R_r}$. The kick arises from the radial flux variation and deformation of the resonance surface, primarily driven by the geometric distortion of the neutrinosphere, rather than any change in a conserved plane-parallel flux due to %EE: Please check that the intended meaning has been retained.
the star’s overall structure~\cite{Barkovich:2002wh}. The fractional anisotropy of momenta in this scenario is given as~\cite{Barkovich:2002wh}
\begin{equation}
 \frac{\delta k}{k}=-\frac{1}{18}\frac{\delta}{R_r},
 \label{eq:3}
\end{equation}
where $R_r$ represents the radius of the resonance surface, while the value of $\delta$ depends on the proto-NS model, such as the polytropic~\cite{Goldreich1980,Raffelt:1996wa,Shapiro:1983du,Weinberg:1972kfs} or spherical Eddington models~\cite{Schinder1982}. However, in this scenario, a magnetic field on the order of $10^{16}~\mathrm{G}$ and a neutrino mass around $100~\mathrm{eV}$ are required.

\subsection{Sterile Neutrino, Dark Matter, and Other New Physics Scenarios}
\subsubsection{Sterile Neutrino and Dark Matter}

The need for a heavy active neutrino mass to explain the pulsar kick, which conflicts with neutrino oscillation and cosmological predictions, can be mitigated by considering active--sterile neutrino oscillations. For active--active oscillations ($\nu_e \leftrightarrow \nu_{\mu, \tau}$), thermalization occurs when incoming $\nu_{\mu, \tau}$ either enter their own neutrinosphere within the proto-NS or reconvert to $\nu_e$ after crossing the resonance region, placing them within their neutrinosphere. In contrast, for sterile neutrinos, only the latter process applies. Outgoing neutrinos alone contribute to the net momentum flux, with asymmetry arising from the differential areas of hemispherical emission surfaces for neutrinos moving in opposite directions with equal momentum, generating a non-zero fractional momentum asymmetry. Explaining pulsar kicks through active--sterile neutrino oscillations requires a sterile neutrino mass around the $\mathrm{keV}$ scale, making it a potential candidate for warm DM, along with a very small active--sterile mixing angle~\cite{Kusenko:2004mm,Barkovich:2004jp}. Additionally, a magnetic field of approximately $10^{17}~\mathrm{G}$ is needed, which could plausibly exist in the interior of a proto-NS.

Introducing ultralight DM in the proto-star environment can reduce the need for extreme magnetic fields~\cite{Lambiase:2023hpq}. The alignment of the DM velocity relative to neutrino momentum in different directions creates an asymmetry that could account for the pulsar kick. The ultralight scalar DM couples with neutrinos through derivative coupling as
\begin{equation}
 -\mathcal{L}\supset g^\prime_{\alpha\beta}\partial_\mu\phi\bar{\nu}_\alpha\gamma^\mu\nu_\beta,     
\end{equation}
where $g^\prime_{\alpha\beta}$ denotes the derivative coupling between DM and active neutrinos, and $\phi$ denotes the oscillating DM scalar field given as $\phi=(\sqrt{2\rho_\mathrm{DM}}/m_\phi)\cos(m_\phi t+\theta)$. Here, $m_\phi$ denotes the mass of the scalar field, $\theta$ denotes a random phase, and $\rho_\mathrm{DM}$ denotes the DM density. Therefore, the dispersion relation for neutrinos becomes 
\begin{equation}
E_\nu=|\mathbf{k}|-g^\prime_{\alpha\beta}\dot{\phi}+g^\prime_{\alpha\beta}\hat{\mathbf{k}}\cdot \nabla\phi+\frac{m^2}{2|\mathbf{k}|}+\mathcal{O}({g^\prime_{\alpha\beta}}^2(\nabla\phi)^2).   
 \label{nw}
\end{equation}

The third term in the right-hand side of Equation~(\ref{nw}) is responsible for the pulsar kick. Note that the scalar coupling of $\phi$ with neutrino does not change the neutrino momentum, and the corresponding coupling will not have any effect on the pulsar kick. 

The resonance condition for active--sterile neutrino oscillations within an ultralight scalar DM background is expressed as ~\cite{Lambiase:2023hpq}
\begin{equation}
\frac{\Delta m^2}{2k}\cos2\theta=V_{\nu}+g^\prime_{\alpha\beta}\sqrt{2\rho_{\mathrm{DM}}}\hat{\mathbf{k}} \cdot \mathbf{v}_\phi, 
\label{eq:4}
\end{equation}
where $V_{\nu}$ represents the standard matter potential, and $\mathbf{v}_\phi$ is the DM velocity vector. The coupling of scalar with sterile neutrinos is assumed to be small. The resulting neutrino momentum asymmetry under a scalar DM background becomes~\cite{Lambiase:2023hpq}
\begin{equation}
\frac{\delta k}{k}=\frac{g^\prime_{\alpha\beta} v_\phi\sqrt{2\rho_{\mathrm{DM}}}}{18 V_\nu R_r(h^{-1}_p+h^{-1}_{V_\nu}) },
\label{eq:5}
\end{equation}
where $h^{-1}_p$ and $h^{-1}_{V_\nu}$ are quantities that depend on the proto-star model, and their sum is given as
\begin{equation}
    h^{-1}_p+h^{-1}_{V_\nu}= -\frac{x_\mathrm{res}}{r_\mathrm{res}}\lambda_\Gamma \eta,
    \label{cheque}
\end{equation}
where 
\begin{equation}
    \lambda_\Gamma=(GM_c/r_c\rho_c^{\Gamma-1})(\Gamma-1)/K\Gamma\simeq 0.29\frac{2GM_c}{\mathrm{km}}\frac{10~\mathrm{km}}{r_c}\frac{40~\mathrm{MeV}}{T_c},
     \label{eos6}
    \end{equation}
 and where $\Gamma=4/3$ and $M_c=1~M_\odot$. Also, 
\begin{equation}
\eta=\frac{2\lambda_c}{\lambda_\Gamma}\epsilon^2 +3(2\mu-1)-6(\mu-1)x_\mathrm{res},    
\end{equation}
where $\epsilon=T_c/T(r_\mathrm{res})$.

The ultralight vector DM can also couple to the neutrinos in supernova as 
\begin{equation}
 -\mathcal{L}\supset g^\prime_{\alpha\beta}\bar{\nu}_\alpha \gamma^\mu \nu_\beta A^\prime_\mu,   
\end{equation}
and the neutrino momentum asymmetry under a vector DM background becomes~\cite{Lambiase:2023hpq}
\begin{equation}
   \frac{\delta k}{k}=\frac{g^\prime_{\alpha\beta} \sqrt{2\rho_{\mathrm{DM}}}}{18 V_\nu R_r m_{A^\prime}(h^{-1}_p+h^{-1}_{V_\nu}) }, 
\end{equation}
where $m_{A^\prime}$ denotes the mass of the ultralight vector. 

Below, we discuss one of the proto-star models, specifically the polytropic EoS.
\begin{itemize}
    \item The polytropic model: %MDPI: Please confirm if the bold is unnecessary and can be removed. The following highlights are the same.
The isotropic neutrinosphere in the polytropic model is described by the following equations~\cite{Goldreich1980,Raffelt:1996wa,Shapiro:1983du,Weinberg:1972kfs}
    \begin{eqnarray}
        \frac{dP(r)}{dr}&=&-\frac{GM(r)\rho(r)}{r^2}, ~~~\mathrm{(hydrodynamical~equilibrium)} \label{eos1}\\
        F(r)&=&-\frac{1}{36}\frac{1}{\kappa\rho(r)}\frac{dT^2}{dr}, \mathrm{(energy~transport)} \label{eos2}\\
        F(r)&=&\frac{L_c}{4\pi r^2},\hspace{1.5cm}\mathrm{(flux~conservation)}\label{eos3}
    \end{eqnarray}
    where $P$ denotes the pressure, $F$ denotes the flux, $G$ denotes Newton's gravitational constant, $M(r)=4\pi \int^r_0 {r^\prime}^2 dr^\prime \rho_T(r^\prime)$, $\rho_T$ is the total density of matter, and $L_c$ denotes the luminosity of the proto-star. When the proto-star is assumed to be filled with relativistic nucleons of polytropic gas, the EoS can be written as
    \begin{equation}
        P(r)=K\rho^\Gamma, \hspace{1cm} K=\frac{T_c}{m_n\rho_c^{1/3}}=5.6\times 10^{-5}~\mathrm{MeV}^{-4/3},
        \label{eos4}
    \end{equation}
    where $\Gamma$, $m_n$, $T_c$, and $\rho_c$ denote the adiabatic index, nucleon mass, temperature, and density of the core, respectively, and $T_c=40~\mathrm{MeV}$, $\rho_c=10^{14}~\mathrm{gm/cm^3}$. Using Equations~(\ref{eos1}) and (\ref{eos4}), we can write
    \begin{equation}
        \frac{d\rho^{\Gamma-1}}{dr}=-\frac{\lambda_\Gamma r_c\rho_c^{\Gamma-1}}{M_c}\frac{M(r)}{r^2},
         \label{eos5}
    \end{equation}
    where $r_c$, $\rho_c$, and $M_c$ denote the radius, density, and mass of the core of the proto-star. The quantity $\lambda_\Gamma$ is given in Equation~(\ref{eos6}).
    
    The approximate solution of Equation~(\ref{eos5}) is obtained as
    \begin{equation}
        \rho^{\Gamma-1}(r)=\rho_c^{\Gamma-1}\Big[\lambda_\Gamma\Big(\frac{r_c}{r}-1\Big)m(r)+1\Big],
     \label{eos7}
    \end{equation}
    where $m(r)=\mu+(1-\mu)r_c/r$ with $m(r_c)=1$. Also, by defining $x=r_c/r$, and $\alpha=(1-\mu)\lambda_\Gamma$, $\beta=(2\mu-1)\lambda_\Gamma$, $\gamma=1-\mu\lambda_\Gamma$, we write Equation~(\ref{eos7}) as 
    \begin{equation}
        \rho^{\Gamma-1}(x)=\rho_c^{\Gamma-1}(\alpha x^2+\beta x+\gamma).
        \label{eos8}
    \end{equation}
    
    The parameter $\mu$ is obtained by setting the condition $\rho(R_s)=0$, where $R_s$ is the radius of the star, and we obtain
    \begin{equation}
        \mu=\Big[\frac{R_s}{\lambda_\Gamma(R_s-r_c)}-\frac{r_c}{R_s}\Big]\frac{R_s}{R_s-r_c}.
         \label{eos9}
    \end{equation}
    
    Also, combining Equations~(\ref{eos2}) and (\ref{eos3}), we can write the expression of the temperature profile in terms of density distribution as
    \begin{equation}
        \frac{dT^2}{dr}=-\frac{9\kappa L_c}{\pi r^2}\rho(r),
         \label{eos10}
    \end{equation}
    where the core luminosity is $L_c\sim 9.5\times 10^{51}~\mathrm{erg/s}$ and $\kappa\sim 5.6\times 10^{-9}\mathrm{cm^4/erg^3s^2}$. The solution of Equation~(\ref{eos10}) is obtained as 
    \begin{equation}
        T(r)=T_c\sqrt{{2\lambda_c}\Big[\chi(r_c/r)-\chi(1)+1\Big]},
        \label{eos11}
    \end{equation}
    where
    \begin{equation}
        \lambda_c=\frac{9}{2\pi}\frac{\kappa L_c\rho_c}{T^2_c r_c}\sim 1.95\frac{\rho_c}{10^{14}~\mathrm{g/cm^3}}\frac{10~\mathrm{km}}{r_c}\Big(\frac{40~\mathrm{MeV}}{T_c}\Big)^2,
        \label{eos12}
    \end{equation}
    and $\chi$ is a polynomial function of $x$, given as
    \begin{equation}
    \begin{split}
        \chi(x)=\gamma^3 x+\frac{3}{2}\beta\gamma^2x^2+\gamma(\alpha\gamma+\beta^2)x^3+\frac{\beta}{4}(6\alpha\gamma+\beta^2)x^4+\\ 
        +\frac{3\alpha}{5}(\alpha\gamma+\beta^2)x^5+\frac{\beta\alpha^2}{2}x^6+
        \frac{\alpha^3}{7}x^7.
        \end{split}
    \end{equation}
\end{itemize}

To achieve a 1\% momentum asymmetry required for the pulsar kick, the coupling and mass parameters are estimated as $g^\prime\sim 1.23\times 10^{-11}~\mathrm{eV}^{-1}$ and $m_\phi\sim 1.02\times 10^{-14}~\mathrm{eV}$. The asymmetric neutrino emission during core collapse induces a lasting strain in the surrounding spacetime, detectable via second- and third-generation GW detectors~\cite{Loveridge:2003fy}. The maximum characteristic GW amplitude from this neutrino emission is given by~\cite{Loveridge:2003fy,Lambiase:2023hpq}
\begin{equation}
|h(t)|\simeq \frac{2G}{r}\alpha L_\nu\mathcal{\tau}\lesssim 1.06\times 10^{-19}\Big(\frac{\alpha}{0.01}\Big)\Big(\frac{E_\mathrm{tot}}{2\times 10^{53}~\mathrm{erg}}\Big)\Big(\frac{1~\mathrm{kpc}}{r}\Big),
\label{eq:6}
\end{equation}
where $\alpha$ represents the anisotropy parameter, $L_\nu$ the neutrino luminosity, and $r$ the distance from the source. The characteristic neutrino emission timescale $\tau$ is defined as $E_{\mathrm{tot}}=L_\nu \tau$.

The anisotropy parameter depends on the DM mass and coupling, as well as alternative DM models, which can also be constrained by pulsar kick observations. Consequently, GW detectors offer a unique avenue for probing DM interactions. 

A comprehensive analysis of $\alpha$ necessitates examining its time evolution during the supernova collapse. GW strain can also be expressed in the frequency domain for both periodic and burst signals. Given that pulsars typically have frequencies of $\mathcal{O}(\mathrm{kHz})$, and bursts last only a few seconds, the source is often treated as periodic. These periodic signals fall within the detection range of LIGO $(10-1000~\mathrm{Hz})$, and the corresponding strain in the frequency domain is expressed as~\cite{Loveridge:2003fy}
\begin{equation}
h(f)\lesssim 10^{-24}~\mathrm{Hz}^{-1/2}\Big(\frac{\alpha}{0.01}\Big)\Big(\frac{10~\mathrm{s}}{\tau}\Big)^{1/2}\Big(\frac{1~\mathrm{kpc}}{r}\Big)\Big(\frac{1~\mathrm{kHz}}{f}\Big).    
\end{equation}

If the signal frequency is low, corresponding to a few pulsar rotations during the neutrino ejection time, with a frequency range of $0.001-0.1~\mathrm{Hz}$, such burst signals can be detected via LISA. In this case, the strain in the frequency domain depends on the anisotropy parameter and is given as~\cite{Loveridge:2003fy}
\begin{equation}
h(f)\lesssim 3\times 10^{-22}~\mathrm{Hz}^{-1/2}\Big(\frac{\alpha}{0.01}\Big)\Big(\frac{10~\mathrm{s}}{\tau}\Big)\Big(\frac{1~\mathrm{kpc}}{r}\Big)\Big(\frac{1~\mathrm{Hz}}{f}\Big)^{3/2}.    
\end{equation}

\begin{figure}[H]
   % \centering
    \includegraphics[width=0.8\linewidth]{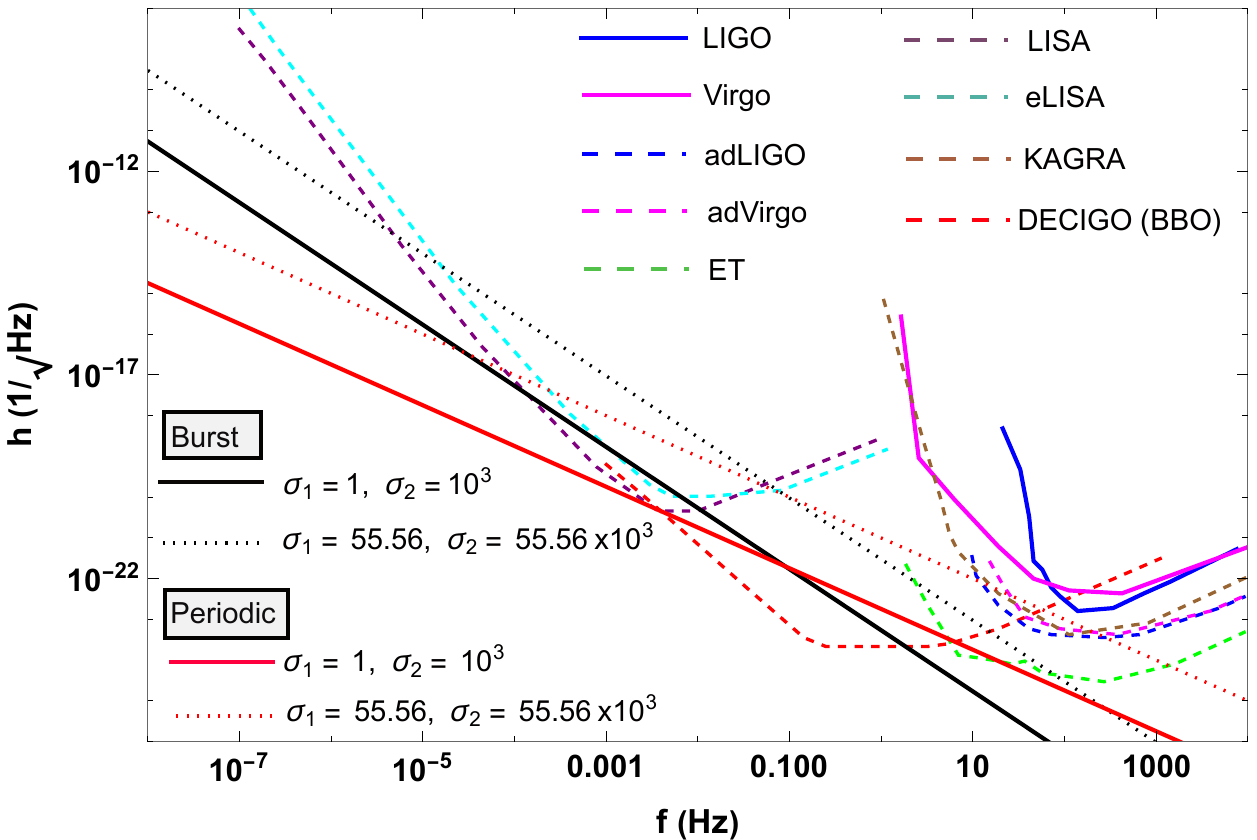}
    \caption{The variation in GW intensity with radiation frequency is presented, illustrating a pulsar kick driven by asymmetric neutrino emission induced via ultralight DM. Sensitivity curves of GW detectors are included for reference. The plot is adapted from~\cite{Lambiase:2023hpq}.}
    \label{memory}
\end{figure}

 Figure \ref{memory} illustrates the variation in GW intensity with frequency, providing insights into pulsar kicks driven by asymmetric neutrino emission induced via ultralight DM, which is adapted from~\cite{Lambiase:2023hpq}. For ultralight vector DM, $\alpha=1.8\times 10^{-4}~\sigma_1$, where $\sigma_1=g^{\prime}_{\alpha\beta}/{m_{A^\prime} x_{\mathrm{res}}\eta}$, and for ultralight scalar DM, $\alpha=1.8\times 10^{-7}~\sigma_2$, where $\sigma_2=g^{\prime}_{\alpha\beta}/{x_\mathrm{res}\eta}$. Here, $x_\mathrm{res}=r_c/r_\mathrm{res}$, and $\eta<1$ is a parameter that depends on the proto-star model. Thus, the couplings and mass of DM can be constrained by analyzing both periodic and burst signals using second- and third-generation GW detectors. The red solid and dotted lines represent our results, illustrating how the DM coupling can be constrained through the periodic signal of the pulsar; these results reflect the relationship between the signal and the constraints we derive. On the other hand, the black solid and dotted lines correspond to our findings related to the constraints on the DM coupling, which are obtained from the burst signal of the pulsar; this approach, too, highlights the sensitivity of the coupling to different observational data, demonstrating how both the periodic and burst signals can offer complementary insights into the properties of DM. The other lines in Figure \ref{memory} correspond to the sensitivity curves of the existing and upcoming GW detectors. The GW intensity decreases with an increasing radiation frequency, while higher ultralight DM-neutrino coupling enhances the wave intensity. Signals from nearby supernovae with appropriate rotation frequencies are expected to be detectable via future GW detectors. The neutrino-induced gravitational memory effect in the presence of ultralight DM, resulting from a supernova, lies below the sensitivity of current detectors like LIGO and Virgo but can be observed via second- and third-generation detectors. Rapidly rotating NSs with larger anisotropy parameters may be detected via advanced detectors such as adLIGO and adVirgo, whereas slowly rotating NSs are within the sensitivity range of LISA. GW detectors operating in the deci-Hz \mbox{range~\cite{Seto:2001qf,TianQin:2015yph,Graham:2016plp,Ruan:2018tsw},} such as DECIGO (DECi-hertz Interferometer Gravitational Wave Observatory)~\cite{Yagi:2011wg} and BBO (Big Bang Observer)~\cite{Yagi:2011wg}, which achieve a sensitivity of GW strain around $10^{-24}$, could detect memory signals from pulsar kicks. Furthermore, the Einstein Telescope (ET)~\cite{Maggiore:2019uih}, with its unprecedented sensitivity ($10^{-24}-10^{-25}$), will explore GW signals from cosmological distances, enabling studies of supernova neutrinos and the associated gravitational memory effects. These advanced detectors will also be capable of probing smaller anisotropy parameters for asymmetric neutrino emissions from sources as distant as a few megaparsecs. Detecting such gravitational memory signals will offer valuable constraints on sterile neutrino masses, ultralight DM interactions, and other beyond SM scenarios.

\subsubsection{Majoron Emission}

If, rather than ultralight DM or sterile neutrinos, active neutrinos couple to a massless pseudoscalar majoron, an asymmetric emission of majorons could account for the pulsar kick, provided that majorons carry away at least $10\%$ of the star's total binding energy~\cite{Farzan:2005yp}. A magnetic field around $10^{16}~\mathrm{G}$ is necessary for this explanation. In such a strong magnetic field, electrons in the medium become polarized, introducing an additional term to the neutrino potential given by~\cite{Farzan:2005yp}
\begin{equation}
\delta V=-\sqrt{2}G_F n_B \langle\lambda_e\rangle \cos\theta_k\Big(\frac{3}{2},\frac{3}{2},\frac{1}{2}\Big),
\label{eq:7}
\end{equation}
where $n_B$ represents the baryon density, $Y_e=(n_e-\bar{n}_e)/n_B$ is the electron fraction, $\theta_k$ is the angle between the neutrino momentum and polarization direction, and $\langle\lambda_e\rangle$ denotes the fraction of polarized electrons, defined as~\cite{Farzan:2005yp}
\begin{equation}
\langle\vec{\lambda}_e\rangle=-\frac{e\vec{B}}{2}\Big(\frac{3}{\pi^4}\Big)^{1/3}n^{-2/3}_e.
\label{eq:8}
\end{equation}

Since $\delta V$ depends on the direction of the neutrino momentum, the majoron emission processes, $\nu\nu\rightarrow\phi$ and $\bar{\nu}\rightarrow \nu\phi$, will vary based on the initial neutrino directions, leading to anisotropic emission. The magnetic field strength necessary to account for the pulsar kick via the process $\nu_e\nu_e\rightarrow \phi$ is given by~\cite{Farzan:2005yp}
\begin{equation}
|\vec{B}|=3\times 10^{16}~\mathrm{G}\frac{M}{1.4~M_\odot}\frac{v}{500~\mathrm{km/s}}\frac{3\times 10^{53}~\mathrm{erg}}{E_{\mathrm{tot}}}\frac{V_\nu}{0.5~\mathrm{eV}}\Big(\frac{0.05~\mathrm{fm}^{-3}}{n_e}\Big)^{1/3}\frac{0.5}{x}.
\label{eq:9}
\end{equation}

\subsubsection{Neutrino Spin--Flavor Oscillation} The pulsar kick mechanism can also be attributed to the spin polarization of matter in a supernova's magnetic field, incorporating neutrino spin--flavor precession. This magnetic field distorts the resonance surface for neutrino spin--flavor transitions, leading to anisotropic neutrino emissions. In cases where resonant spin--flavor precession occurs in sterile neutrino states, the magnetic field strength required to produce the pulsar kick is approximately $B \gtrsim 4 \times 10^{15}~\mathrm{G}$, assuming a Dirac transition magnetic moment of $\mu_\nu \gtrsim 10^{-14}~\mu_B$~\cite{Akhmedov:1997qb}. Likewise, for resonant spin--flavor precession between active neutrinos and anti-neutrinos, a magnetic field of $B \gtrsim 2 \times 10^{16}~\mathrm{G}$ is needed, with a Majorana transition magnetic moment of $\mu_\nu \gtrsim 10^{-15}~\mu_B$.

The chirality flip of neutrinos, driven by the presence of a neutrino magnetic moment in strange quark matter NSs, offers an explanation for the pulsar kick. For this mechanism to account for the average pulsar velocity, the required neutrino magnetic moment is approximately $\mu_\nu \sim$ 3.6 $\times 10^{-18}~\mu_B$~\cite{Ayala:2024wgb}.

The spin--gravity coupling of neutrinos in the gravitational field of a rotating NS can lead to a pulsar kick.
When the magnetic and gravitational interactions are taken into account, the spin flavor conversion for neutrinos is governed by the evolution equation~\cite{Lambiase:2004qk}
\begin{equation}
\iota \frac{d}{d\lambda}\begin{pmatrix}
    \nu_{fL}\\\nu_{f^\prime L}\\\nu_{fR}\\\nu_{f^\prime R} 
    \end{pmatrix}
    =H\begin{pmatrix}
        \nu_{fL}\\\nu_{f^\prime L}\\\nu_{fR}\\\nu_{f^\prime R}
    \end{pmatrix},
    \label{eq:10}
\end{equation}
where the effective Hamiltonian on the chiral basis can be written as~\cite{Lambiase:2004qk}
\begin{equation}
H=
    \begin{pmatrix}
        H_L & H^*_{ff^\prime}\\
        H_{ff^\prime} & H_R
    \end{pmatrix},
    \label{eq:11}
\end{equation}
and 
\begin{equation}
H_L=\begin{pmatrix}
    V_{\nu_f}+\Omega_G-\delta c_2 & \delta s_2\\
    \delta s_2 &  V_{\nu_{f^\prime}}+\Omega_G+\delta c_2
\end{pmatrix}\,,
\end{equation}
\begin{equation}
H_R=\begin{pmatrix}
-\delta c_2 & \delta s_2\\
\delta s_2 & \delta c_2
\end{pmatrix}, \quad 
H_{ff^\prime}=B_\perp\begin{pmatrix}
    \mu_{ff}& \mu_{ff^\prime}\\
    \mu_{ff^\prime} & \mu_{f^\prime f^\prime}
\end{pmatrix},
 \label{eq:12}
\end{equation}
where $c_2=\cos2\theta$, $s_2=\sin2\theta$, $\delta=(m^2_2-m^2_1)/4E$, $m_{1,2}$ are the two neutrino mass eigenstates, $B_\perp=B\sin\alpha$ is the component of the magnetic field orthogonal to neutrino momentum, and~\cite{Lambiase:2004qk}
\begin{equation}
    \Omega_G (r)=8\times 10^{-13} \Big(\frac{M}{M_\odot}\Big)\Big(\frac{R}{10~\mathrm{km}}\Big)^2\Big(\frac{10~\mathrm{km}}{r}\Big)^3 \frac{\omega^\prime \cos\beta}{10^4~\mathrm{Hz}}~\mathrm{eV},
     \label{eq:13}
\end{equation}
where $\omega^\prime$ is related to the angular velocity of NS, and $\beta$ is the angle between the neutrino momentum and the angular velocity of NS. The anisotropy in neutrino emission arises from the alignment of neutrino momenta relative to the star's angular velocity. The fraction of pulsar momentum asymmetry in this scenario is given as~\cite{Lambiase:2004qk}
\begin{equation}
    \frac{\delta k}{k}=\frac{4GMR^2\omega^\prime}{5r^3}\frac{1}{9V_{\nu_f}R_r(h^{-1}_p+h^{-1}_{V_{\nu_f}})}.
    \label{eq:14}
\end{equation}

In this scenario, spin--flavor oscillation is considered for calculating the momentum asymmetry of the pulsar, with a required neutrino magnetic moment of approximately $\mu_\nu \lesssim 10^{-11}~\mu_B$ to account for the kick.

\subsubsection{Lorentz and CPT Violation}Pulsar kick can also be explained from the active--sterile neutrino oscillation in the Standard Model Extension (SME) framework, where the Lorentz and/or CPT invariance can be violated. The neutrino dispersion relation in such framework can be written as~\cite{Lambiase:2005iz}
\begin{equation}
 E\simeq k+\frac{m^2}{2k}+\Omega,   
\label{eq:15}
\end{equation}
where
\begin{equation}
\Omega=-\frac{c_L^{\mu\nu}k_\mu k_\nu}{k}+\frac{a_{L\mu}k^\mu}{k}, 
\label{eq:16}
\end{equation}
where $c$ violates Lorentz invariance, and $a$ violates both CPT and Lorentz invariance. The resonance condition for the active--sterile neutrino oscillation in this SME framework can be written as ~\cite{Lambiase:2005iz} %MDPI: \hl{} %MDPI: Please confirm if the bold is unnecessary and can be removed from text and equations.
\begin{equation}
2\delta c_2=V_{\nu_f}+\delta\mathbf{\Pi}\cdot{\hat{\mathbf{k}}}, 
\label{eq:17}
\end{equation}
where $\delta\mathbf{\Pi}\cdot\hat{\mathbf{k}}=(\delta \mathbf{a}-2 k\delta \mathbf{c})\cdot \hat{\mathbf{k}}$ and $\delta \mathbf{c}=\mathbf{c}^{(\nu_f)}-\mathbf{c}^{(\nu_s)}$, $\delta \mathbf{a}=\mathbf{a}^{(\nu_f)}-\mathbf{a}^{(\nu_s)}$. Therefore, the fractional asymmetry in the neutrino momenta can be written as~\cite{Lambiase:2005iz}
\begin{equation}
\frac{\delta k}{k}=(2k|\delta \mathbf{c_L}|+|\delta \mathbf{a_L}|)\frac{1}{9 R_r V_{\nu_f}(h^{-1}_p+h^{-1}_{V_{\nu_f}})}.
\label{eq:18}
\end{equation}

For $1\%$ of momentum asymmetry of a pulsar, the SME parameters can be constrained as $\delta{\mathbf{c}_L}\lesssim 4\times 10^{-17}$ and $\delta{\mathbf{a}_L}\lesssim 1.6\times 10^{-9}~\mathrm{eV}$. The SME parameters and the ultralight scalar DM parameters are related in the pulsar kick scenario as $a_L=g^\prime|\nabla\phi|$ and $c_L=g^\prime |\nabla\phi|/2k$.

\subsubsection{Massless Neutrino Framework}A pulsar kick can be explained even with massless neutrinos if a strong magnetic field is present, assuming a slight violation of weak universality. In this scenario, one does not need any neutrino magnetic moment. 

The observed pulsar velocity can also be attributed to tensor neutrino--gravity coupling, which causes anisotropic neutrino emission under strong magnetic fields if neutrinos violate the equivalence principle at a level of $10^{-9}-10^{-10}$~\cite{Horvat:1998st}. In scenarios where neutrinos interact with the gradient of the gravitational potential (as can occur in string theory), the dimensional parameter characterizing this violation of the equivalence principle (VEP) needed to account for the pulsar kick is around $(10^{-2}-10^{-3})~\mathrm{cm}$.

Neutrino oscillations induced via VEP, based on the parametrized post-Newtonian (PPN) framework, can account for the pulsar kick~\cite{Barkovich:2001rp}. In this scenario, VEP effects generate anisotropies in both the linear and angular momenta of emitted neutrinos, potentially explaining both the translational and rotational motions observed in pulsars. The massless neutrino dispersion relation under these conditions is given by~\cite{Casini:1998ur}
\begin{equation}
E=k\Big[1+h_{oi}\hat{\mathbf{k}}_i-\frac{1}{2}h_{ij}\hat{\mathbf{k}}_i\hat{\mathbf{k}}_j-\frac{1}{2}h_{00}U\Big],
\label{eq:19}
\end{equation}
where 
\begin{equation}
   h_{00}=2\gamma^\prime U+\mathcal{O}(\omega^4),
\end{equation}
\begin{equation}
   h_{0i}=-\frac{7}{2}\Delta_1 V_i-\frac{1}{2}\Delta_2 W_i+\Big(\alpha_2-\frac{1}{2}\alpha_1\Big)v_i U-\alpha_2 v_jU_{ji}+\mathcal{O}(\omega^4),
\end{equation}
\begin{equation}
h_{ij}=2\gamma U\delta_{ij}+\Gamma U_{ij}+\mathcal{O}(\omega^4).
\end{equation}

Here, $\gamma$, $\gamma^\prime$, $\Gamma$, $\Delta_{1,2}$, $\alpha_{1,2}$, and $\mathbf{v}$ denote the PPN expansion parameters. Also, $\omega$ denotes the velocity of the source. The parameters $\alpha_{1,2}$ vanish in the Lorentz covariant theories; however, if there exists a preferred frame characterized by velocity $\mathbf{v}$, then $\alpha_{1,2}$ are non-zero. Also, $U$, $V_i$, $W_{i}$, and $U_{ji}$ are the potential terms. In this scenario, the neutrino evolution equation can be written as~\cite{Barkovich:2001rp}
\begin{equation}
    \iota \frac{d}{dr}\begin{pmatrix}
        \nu_e \\ \nu_\mu
    \end{pmatrix}
    =\frac{\Delta_0}{2}\begin{pmatrix}
        -\cos2\theta_g & \sin 2\theta_g\\
        \sin2\theta_g &\cos 2\theta_g
    \end{pmatrix}
    \begin{pmatrix}
        \nu_e \\ \nu_\mu
    \end{pmatrix},
\end{equation}
where $\theta_g$ denotes the mixing angle between the flavor and gravitational eigenstate, and $\Delta_0$ for a rotating proto-NS can be written as~\cite{Barkovich:2001rp}
\begin{equation}
\begin{split}
\Delta_0=\Big[-(\delta\gamma^\prime+\delta\gamma)U-\delta\Gamma J-\delta\Gamma I(\hat{\mathbf{r}}\cdot\hat{\mathbf{k}})^2+\Big\{(\delta\alpha_2-\frac{1}{2}\delta\alpha_1)U-\delta\alpha_2 J\Big\}\mathbf{v}\cdot{\hat{\mathbf{k}}}-\\
\delta\alpha_2 I(\hat{\mathbf{r}}\cdot\mathbf{v})\hat{\mathbf{r}}\cdot \hat{\mathbf{k}}-\frac{1}{2}(7\delta\Delta_1+\delta\Delta_2)J\mathbf{\Omega}\times \mathbf{r}\cdot\hat{\mathbf{k}}\Big]E,    
\end{split}
\end{equation}
where $\delta\gamma=\gamma^2-\gamma^1$, $I$, and $J$ are related to the potentials, and $\Delta_0$ plays the same role as $(m^2_2-m^2_1)/4E$.
\subsection{Hydrodynamic Instabilities}\label{sec2.3}
The globally anisotropic ejection of mass and subsequent NS acceleration in a supernova core can result from convective overturn and low-order oscillations in the neutrino-heated layer, which is called the standing accretion shock instability (SASI). Specifically, low-mode $(l=1, 2)$ convection arises from initial random perturbations, particularly when the explosion onset is gradual, allowing convective structures to merge~\cite{Scheck:2006rw}. This kick mechanism does not strictly align the pulsar's spin with its spatial velocity, although the rotation axis serves as a natural orientation, potentially limiting extreme misalignment. Both gravitational and hydrodynamic forces from the asymmetric supernova ejecta propel the NS over a timescale exceeding one second, achieving speeds above $500~\mathrm{km/s}$.

\begin{figure}[H]
   % \centering
    \includegraphics[width=0.8\linewidth]{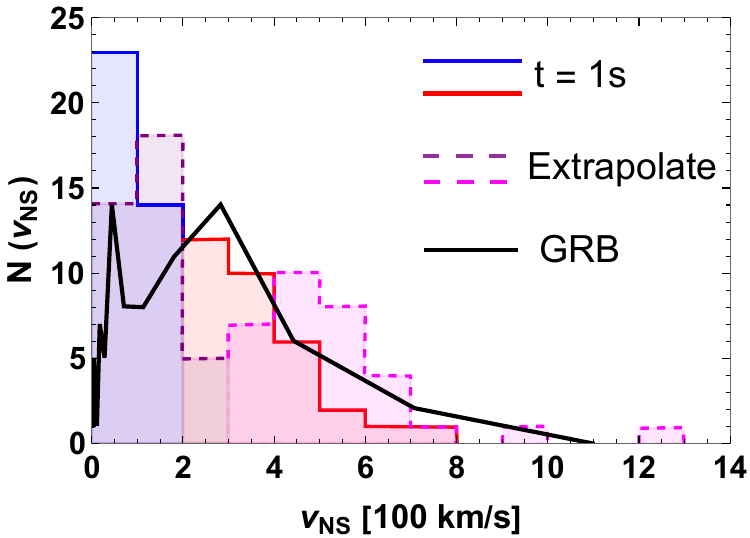}
    \caption{The histograms display the velocity distribution of NSs for the $70$ models detailed in \mbox{Tables A.1–A.5 of~\cite{Scheck:2006rw}}. The solid blue and red line represents the velocity distribution at $t=1~\mathrm{s}$, and the dashed purple and magneta lines correspond to the extrapolation. The red-shaded region indicates the subset of models where NSs attain velocities exceeding $200~\mathrm{km/s}$ within one second of the core bounce. The black solid line correspond to the modeled kick distribution from observed fluence and redshift of gamma ray bursts (GRBs)~\cite{Cui:2007qb}. The figure is adapted from~\cite{Scheck:2006rw,Cui:2007qb}.}
    \label{p1}
\end{figure}

Figure \ref{p1} reveals that NSs in the high-velocity, high-acceleration group (red-shaded region) continue to accelerate to much higher velocities beyond one second, whereas only a few low-velocity stars exceed 200 km/s (blue-shaded region)~\cite{Scheck:2006rw}. After one second of post-bounce evolution, simulated for most models, NS velocities reach up to 800 km/s. The models split into two distinct populations: one with velocities below 200 km/s and low acceleration, and another with velocities above 200 km/s, exhibiting higher average acceleration. The purple and magenta shaded regions correspond to the low and high acceleration regions, respectively, when the NS velocity is extrapolated to its terminal value. The black solid line represents the modeled kick distribution derived from the observed fluence and redshift of GRBs~\cite{Cui:2007qb}. This distribution aligns closely with the observed pulsar kick distribution, suggesting a potential connection between GRBs and supernovae. The asymmetry in GRBs associated with supernovae may play a role in generating pulsar kicks.

The current analysis has limitations, as it relies on simulations assuming axial symmetry, restricting fluid flow along the polar axis. It remains uncertain how prominent $l=1$ modes of ejecta and sustained matter downflows to the neutron star develop in a fully 3D environment or how common they are. Although initial 3D simulations look promising, detailed results are pending. Additionally, the distribution of NS recoil velocities from 3D models is still unclear. The long simulation times required by the stochastic nature of the proposed hydrodynamic kick mechanism make comprehensive studies computationally infeasible for now, leaving the findings suggestive but not conclusive. 

Hydrodynamic recoil resulting from neutrino-driven CCSNe provides a straightforward mechanism for accelerating NSs and pulsars, eliminating the need for assumptions of anisotropic neutrino emission or other exotic models. In a two-dimensional radiation-hydrodynamic simulation, the collapse leads to substantial acceleration of a newly formed NS as a result of a neutrino-driven explosion. During this explosion, a $10\%$ anisotropy in the ejected low-mass, high-velocity material causes a recoil in the high-mass NS, allowing the pulsar to reach recoil speeds of $150~\mathrm{km/s}$. This mechanism offers a consistent explanation for pulsar velocities based on the dynamics of supernova ejecta~\cite{Nordhaus:2010ub}.

The asymmetry of ejecta in supernova remnants (SNRs) provides a possible explanation for the observed pulsar kicks~\cite{Holland-Ashford:2017tqz}. Researchers have studied the connection between NS kick velocities and the X-ray morphologies of 18 SNRs using data from the Chandra X-ray Observatory and ROSAT. They analyzed SNR asymmetries through a power-ratio method, which is a multipole expansion technique that focuses on dipole, quadrupole, and octupole power ratios to quantify asymmetries. The results showed no significant correlation between the power ratio magnitudes and NS kick velocities overall. However, for cases like Cas A and G292.0+1.8, where the X-ray emissions primarily represent ejecta distribution, the associated NSs tended to move in the opposite direction to the bulk of X-ray emissions. Similarly, PKS 1209–51, CTB 109, and Puppis A also exhibited this trend, but since their emissions are dominated by circumstellar or interstellar material, these asymmetries may not directly reflect the distribution of the original supernova ejecta.

Figure 2 of~\cite{Holland-Ashford:2017tqz} presents the quadrupole power ratio ($P_2/P_0$, left) and octupole power ratio ($P_3/P_0$, right) plotted against NS transverse velocities, using the center of emission as the origin for the multipole expansion~\cite{Holland-Ashford:2017tqz}. No significant correlation is observed between $P_2/P_0$ or $P_3/P_0$ and NS velocities. This finding remains consistent for both proper motion estimates and geometric offset velocity estimates, even when accounting for the considerable uncertainties in the geometric measurements. Additionally, neither non-interacting nor interacting SNRs, represented as circles and squares, respectively, in \mbox{Figure 2 of~\cite{Holland-Ashford:2017tqz}}, exhibit a clear trend in power ratios relative to NS velocity. However, it is noteworthy that most SNRs with high $P_2/P_0$ and $P_3/P_0$ values are associated with interacting remnants.
  
The findings support the hypothesis that NS kicks are more likely due to asymmetries in ejecta, rather than anisotropic neutrino emissions. In summary, no relationship was found between NS velocities and power ratio magnitudes within the 0.5–2.1 keV X-ray band, regardless of whether the center of emission or explosion site was chosen as the reference point for the multipole expansions. This indicates that the small-scale X-ray asymmetries and the NS kicks may result from distinct processes or be influenced by different factors.

The simulation also explored the possibility of explaining high recoil velocities through a neutrino-driven supernova explosion mechanism, where non-radial hydrodynamic instabilities break symmetry and induce strong kicks. However, these studies were based on two-dimensional simulations that assume axial symmetry, with the polar axis treated as a coordinate singularity, restricting fluid flow across it.

Hydrodynamical simulations offer valuable insights into pulsar kicks, yet they have limitations. High-resolution, three-dimensional simulations of hydrodynamics and neutrino transport are computationally intensive, which restricts accuracy. While simulations can account for pulsar speeds of a few hundred km/s, achieving velocities above $1000~\mathrm{km/s}$ remains challenging and may require magnetic or neutrino-driven anisotropies. Additionally, the results are highly sensitive to initial conditions, such as the progenitor’s rotation and magnetic field configuration, which influence large-scale asymmetries.
\subsection{Other Mechanisms}
\begin{itemize}
\item Chiral anisotropy conversion: %MDPI: 
Pulsar kicks can be accounted for by the anisotropic emission of neutrinos, which arises from their scattering with the background axial electron current—a result of the chiral separation effect~\cite{Fukushima:2024cpg}. Achieving a pulsar recoil requires anisotropy in either the magnetic field or density in momentum space. In this framework, a magnetic field strength of approximately $10^{16}~\mathrm{G}$ can drive pulsar velocities exceeding $1000~\mathrm{km/s}$.
\item Evanescent proto-neutron star: %MDPI: 
If a CCSN results in a rapidly rotating proto-NS, it may subsequently cool and undergo fragmentation, forming a binary proto-NS system in a very close orbit~\cite{Colpi:2002cu}. In this scenario, the lighter companion could eventually be tidally disrupted, imparting a significant kick to the remaining proto-NS. This mechanism has the potential to generate kick velocities exceeding $1000~\mathrm{km/s}$.
\item Rocket effect: %MDPI: 
An asymmetry in the magnetic field configuration of a pulsar's strong magnetic field could generate a small, continuous electromagnetic force. If the magnetic moment is misaligned with the pulsar's rotation axis, this could lead to a "rocket effect," exerting a gradual push on the pulsar~\cite{Agalianou:2023lvv}. However, a significant drawback of this mechanism is that it cannot achieve kick velocities as high as those produced by neutrino-driven or hydrodynamic processes.
\end{itemize}

\section{Future Outlook}\label{outlook}
\begin{itemize}
\item Modified gravity: %MDPI: 
The physics of NS and compact objects has also been extensively studied in the framework of theories that generalized or modify the GR (see, for example~\cite{Cui:2024nkr, Astashenok:2013vza, Astashenok:2014pua}, and for a review of extended theories of gravity, see~\cite{Capozziello:2011et}). Pulsar kick can be studied in these alternative theories of gravity.

\item Modeling of proto-NS: %MDPI: 
 Within the framework of relativistic mean field (RMF) theory, various constituents of SM or BSM particles can be incorporated, along with the effects of phase transitions~\cite{Maruyama:2012hf}. These modifications to the proto-NS EoS can be explored in detail to analyze their impact on pulsar kicks.

\item Modeling of magnetic field: %MDPI: 
The magnetic field of the proto-NS remains uncertain due to significant observational uncertainties. Nevertheless, various models have been proposed, and studying pulsar kicks within the context of these different models would be highly intriguing.

\item Advanced simulations: %MDPI: 
More advanced simulations, supported by improved computational resources, will provide deeper insights into the mechanisms underlying pulsar kicks.

\item New neutrino interaction: %MDPI: 
In addition to the neutrino interactions discussed earlier, exploring non-standard neutrino interactions, as well as the effects of neutrino decay and their interactions with DM, would be valuable. Conducting such studies with precise timing analyses could offer significant insights into the mechanisms of pulsar~kicks.
\end{itemize}
\section{Conclusions and Discussions}\label{sec2}
The velocity distribution of pulsars is strongly influenced by the internal structure of the proto-NS. Various EoS have been proposed to model the proto-star's evolution, including the polytropic model, the spherical Eddington model, and the Murnaghan EoS~\cite{doi:10.1073/pnas.30.9.244}. These equations describe the relationship between pressure, density, and temperature within the star. However, each of these models has its own inherent limitations and~assumptions.

The theory that pulsar kicks arise from asymmetric neutrino emission does not predict a direct correlation between velocity and a magnetic field (\emph{B}--\emph{V} correlation) because the magnetic field relevant to the kick is located inside the hot NS during the initial seconds after the supernova collapse. Astronomical observations, however, can only infer the surface magnetic field of NSs millions of years after this event. The magnetic fields at early and late stages of NS evolution are not straightforwardly connected due to the complex changes that occur during the star’s cooling process. Future X-ray observatories like XRISM will have the capability to detect weak emission lines from trace elements, potentially providing new insights into linking pulsar kick theories with observational evidence. For example, the magnetic field at the central region has been considered to be $10^{18}~\mathrm{G}$. The magnetic field of the proto-NS depends on the baryon density and its parametrized \mbox{as~\cite{Chandrasekhar:1953zz,Bandyopadhyay:1997kh,Manka:2002pi}}
\begin{equation}
 B_{\mathrm{PNS}}=B_s+B_c[1-e^{-\beta(\rho/\rho_s)^\gamma}],   
\end{equation}
where $B_s= 10^{14}~\mathrm{G}$, $B_c=10^{18}~\mathrm{G}$ are the magnetic fields at the surface and the core, respectively, $\beta=10^{-5}$, $\gamma=3$, and $\rho_s=10^{11}~\mathrm{g/cm^3}$. Thus, the requirement of a strong magnetic field to drive neutrino-induced kicks is not completely ruled out. 

Since no strong correlation exists between pulsar velocity and other pulsar properties, statistical analyses of pulsar populations neither confirm nor disprove any existing models in the literature. Most of the proposed mechanism for pulsar kicks does not depend on a correlation between pulsar velocity and a magnetic field. This result aligns with recent studies suggesting that a B--V correlation is unnecessary to explain pulsar kicks~\cite{Wex:1999kr,Hansen:1997zw,Toscano:1998iz}.

One benefit of explaining pulsar kicks through the hypothesis of massless neutrinos and equivalence principle violation is that it avoids conflicts with cosmological neutrino constraints, unlike the model involving active neutrino oscillations proposed in~\cite{Kusenko:1996sr}.

While establishing a direct relationship between the magnetic field and pulsar velocity is challenging, the duration of neutrino or majoron emission significantly exceeds the initial rotation period of a newborn pulsar. This extended emission period suggests that there may be an alignment between the rotation axis and the direction of the pulsar kick~\cite{Spruit:1998sg,Ng:2003fm,Romani:2004dx}. Although only a few pulsars have known rotational axes that appear to align with their kick directions, this alignment trend hints at a potential correlation. However, further data are essential to reach a definitive conclusion on this connection.

It remains uncertain to what extent pronounced $l=1$ modes in the ejecta distribution and sustained downflows of matter onto the NS can develop in a fully 3D environment, or how frequently these features might occur, though initial 3D simulations using the current setup and input physics show promise. Additionally, the expected distribution of NS recoil velocities from 3D models is still unclear. The hydrodynamic kick mechanism is inherently stochastic and requires extended simulation durations, but the vast computational resources needed make such large-scale, long-term simulations currently infeasible.

Studies~\cite{Wex:1999kr,Spruit:1998sg,Lai:2000pk,Pavlov:2001kx,Helfand:2000qt,Colpi:2002cu,Ng:2007aw} suggest that low-velocity NSs tend to have space velocities aligned with their spin axes, while high-velocity NSs move primarily perpendicularly to their spins, resulting in two distinct (low- and high-velocity) distributions.

This review has addressed the longstanding challenge of explaining pulsar kicks in astrophysics, along with potential solutions to this phenomenon. Pulsars receive a high-velocity kick shortly after their formation in supernova explosions, with birth speeds ranging from several hundred to thousands of kilometers per second, often placing them far from their origin. The three leading explanations for pulsar kicks are asymmetric supernova explosions, asymmetric neutrino emissions, and hydrodynamic instabilities. However, no single mechanism has fully accounted for the observed pulsar velocity distribution. It is likely that a combination of these mechanisms contributes in varying degrees, depending on the specific conditions of each supernova. Advancements in supernova observations, more refined simulations, and deeper insights into neutrino physics in supernova environment may improve our understanding of these powerful kicks. Additionally, studying pulsar kicks provides a valuable opportunity to constrain potential new physics scenarios.
 %%%%%%%%%%%%%%%%%%%%%%%%%%%%%%%%%%%%%%%%%%
%%%%%%%%%%%%%%%%%%%%%%%%%%%%%%%%%%%%%%%%%%
\vspace{6pt} 

%%%%%%%%%%%%%%%%%%%%%%%%%%%%%%%%%%%%%%%%%%
%% optional
%\supplementary{The following supporting information can be downloaded at:  \linksupplementary{s1}, Figure S1: title; Table S1: title; Video S1: title.}


%\begin{tabularx}{\textwidth}{lll}
%\toprule
%\textbf{Name} & \textbf{Type} & \textbf{Description} \\
%\midrule
%S1 & Python script (.py) & Script of python source code used in XX \\
%S2 & Text (.txt) & Script of modelling code used to make Figure X \\
%S3 & Text (.txt) & Raw data from experiment X \\
%S4 & Video (.mp4) & Video demonstrating the hardware in use \\
%\bottomrule
%\end{tabularx}
%}

%%%%%%%%%%%%%%%%%%%%%%%%%%%%%%%%%%%%%%%%%%
\authorcontributions{Conceptualization,
G.L. and T.K.P.; methodology, G.L. and T.K.P.; software, T.K.P.; validation, G.L. and T.K.P.; formal analysis, G.L. and T.K.P.; investigation, G.L. and T.K.P.; resources, G.L. and T.K.P.;
data curation, G.L. and T.K.P.; writing—original draft preparation, G.L. and T.K.P.;
writing—review and editing, G.L. and T.K.P.; visualization, G.L. and T.K.P.; supervision,
G.L.; project administration, G.L. All
authors have read and agreed to the published version of the
manuscript.}

\funding{This research received no external funding.}

\dataavailability{The review article has no associated data.} 

% Only for journal Nursing Reports
%\publicinvolvement{Please describe how the public (patients, consumers, carers) were involved in the research. Consider reporting against the GRIPP2 (Guidance for Reporting Involvement of Patients and the Public) checklist. If the public were not involved in any aspect of the research add: ``No public involvement in any aspect of this research''.}

% Only for journal Nursing Reports
%\guidelinesstandards{Please add a statement indicating which reporting guideline was used when drafting the report. For example, ``This manuscript was drafted against the XXX (the full name of reporting guidelines and citation) for XXX (type of research) research''. A complete list of reporting guidelines can be accessed via the equator network: \url{https://www.equator-network.org/}.}

% Only for journal Nursing Reports
%\useofartificialintelligence{Please describe in detail any and all uses of artificial intelligence (AI) or AI-assisted tools used in the preparation of the manuscript. This may include, but is not limited to, language translation, language editing and grammar, or generating text. Alternatively, please state that “AI or AI-assisted tools were not used in drafting any aspect of this manuscript”.}

\acknowledgments{ G.L. and T.K.P. thank COST Actions COSMIC WISPers CA21106 and BridgeQG CA23130, supported by COST (European Cooperation in Science and Technology).}

\conflictsofinterest{The authors declare no conflicts of interest.} 

%%%%%%%%%%%%%%%%%%%%%%%%%%%%%%%%%%%%%%%%%%
%% Optional

%% Only for journal Encyclopedia
%\entrylink{The Link to this entry published on the encyclopedia platform.}

%%%%%%%%%%%%%%%%%%%%%%%%%%%%%%%%%%%%%%%%%%
%% Optional
\begin{adjustwidth}{-\extralength}{0cm}

\reftitle{References}

\PublishersNote{}
\end{adjustwidth}
\end{document}